\documentclass[pra,aps,twocolumn,showpacs,amsmath,amssymb,floatfix]{revtex4}

\usepackage[dvips]{graphicx} 
\usepackage{bm} 

\newcommand{\be}{\begin{equation}}
\newcommand{\ee}{\end{equation}}
\newcommand{\bea}{\begin{eqnarray}}
\newcommand{\eea}{\end{eqnarray}}
\newcommand{\eqname}[1]{\label{eq:#1}}
\newcommand{\eq}[1]{(\ref{eq:#1})}
\newcommand{\as}{\alpha_\sigma}
\newcommand{\bs}{\beta_\sigma}
\newcommand{\ks}{\chi_\sigma}
\newcommand{\xs}{\xi_\sigma}
\newcommand{\oit}{\omega_i(t)}
\newcommand{\oi}{\omega_i}
\newcommand{\ai}{a_i(t)}

\begin{document}

\title{Spin oscillations of the normal polarized Fermi gas at Unitarity}
\author{Alessio Recati and Sandro Stringari}
\affiliation{Dipartimento di Fisica, Universit\`a di Trento and CNR-INO BEC Center, I-38123 Povo, Italy}

\begin{abstract} 
Using density functional theory in a time dependent approach we determine the frequencies of the compressional modes of the normal phase of a Fermi gas at unitarity as a function of its polarization. Our energy functional accounts for the typical elastic deformations exhibited by Landau theory of Fermi liquids. The comparison with the available experiments is biased by important collisional effects affecting both the {\it in phase} and the {\it out of phase} oscillations even at the lowest temperatures. New experiments in the collisionless regime would provide a crucial test of the applicability of Landau theory to the dynamics of these strongly interacting normal Fermi gases.
\end{abstract} 
 
\maketitle

It is nowadays well accepted that a spin-polarized sample of ultracold Fermi gases at unitarity, where the scattering length becomes  infinite,  is not superfluid, but rather a normal Fermi liquid provided the polarization is large enough (see, e.g.,\cite{importance} and the very recent experiment \cite{ENSeos} and reference therein).  The transition between the superfluid and normal phase occurs at the value $(n_\downarrow/n_\uparrow)_{CC} \sim 0.45$  of the relative concentration between the minority-$\downarrow$ and the majority-$\uparrow$ component. Such a value is also known as the Chandrasekhar-Clogston limit. 

In an harmonically trapped geometry the critical value $(n_\downarrow/n_\uparrow)_{CC}$ is reached in the center of the trap at the value $P=(N_\uparrow-N_\downarrow)/(N_\uparrow+N_\downarrow)\simeq0.77$ of the (global) polarization of the gas, where $N_\uparrow$ and $N_\downarrow$ are the paricle numbers of the majority and of the minority component, respectively.  For smaller values of $P$ we have a superfluid core in the center of the trap, surrounded by a normal shell. In the superfluid phase the densities are equal, at the border they present a jump, just beyond the border they have a ratio equal to the Chandrasekhar-Clogston limit and such a ratio decreases going outward in the trap. For larger value of the polarization no superfluid is instead present in the trap. The occurrence of this behavior is well confirmed experimentally \cite{MITnormal,ENScoll,ENSthermo,ENSeos} and the comparison with the theory of the normal phase, first introduced in \cite{Loboetal}, is remarkably accurate concerning the critical value of $P$ and the density profiles\cite{importance}.

The theory developed in \cite{importance,Loboetal} is based on the assumption that the normal phase can be thought as a Fermi gas of polarons. A polaron in this context is just a single $\downarrow$-impurity immersed in an ideal Fermi sea of $\uparrow$-atoms, which behaves as a quasi-particle with an effective mass $m^*$ and chemical potential conventionally written as $\mu_\downarrow=-3/5A \mu_\uparrow$, with $A>0$ since the interaction between the two species is attractive. The values of the dimensionless parameters $A$ and $m^*/m$ have been calculated by means of different theoretical approaches \cite{Loboetal,chevy,roland,prok,pilati} and first measurements have been reported in \cite{martin,ENScoll,ENSthermo,ENSeos}.

In the presence of an harmonic trapping $V_{ho}({\bf x})=1/2 m(\omega_x^2 x^2+\omega_y^2 y^2+\omega_z^2 z^2)$, the polaron is well described by the single-particle effective Hamiltonian \cite{importance}
\be
H_{sp}=\frac{p^2}{2m^*}+V_{ho}({\bf x})\left(1+{3 \over 5}A\right).
\eqname{effH}
\ee
showing that the impurity feels the renormalized oscillator frequencies 
\be
\tilde\omega_i=\omega_i\;\sqrt{\left(1+\frac{3}{5} A\right)\frac{m}{m^*}} .\eqname{single}
\ee 
Accordingly it will oscillate with frequencies renormalized by same factor $\sqrt{\left(1+3/5 A\right)m/m^*}$.  When more impurities are present their dynamic behaviour will be  affected in a different way by the interaction with the majority component and the frequency of the oscillations will depend on the value of the polarization.

Very recently the results of such an experiment have been reported in \cite{ENScoll}, where the frequencies of the axial breathing modes for the majority and the minority components have been measured at different values of $P$. Intuitively one expects that there are both {\it in phase} and {\it out of phase} modes of the two spin components, the latter corresponding, in the limit of zero concentration, to the polaron oscillation. Indeed in the experiment \cite{ENScoll} it is seen that the majority and the minority component share a common frequency (in phase mode), but the minority one presents also a second branch at higher frequency. 

The aim of the present work is to study the collective mode frequencies of the unitary normal phase as a function of the polarization. To this purpose we develop the Landau formalism of quasi-particles to describe the collisionless regime of the polarized Fermi gas.

A first important issue is the problem of the damping of the oscillation, since collisions can be very effective in this strongly interacting system.  Let us consider highly polarized Fermi gases at a temperature $T$ smaller than the degeneracy temperatures $T_{{\rm F}\sigma}$ of both the components.  
The damping of the dipole oscillation was calculated in \cite{bruunrecati} and  can be written as
\be
\frac{1}{\omega_0 \tau_P}\propto
A^2 (6N_\uparrow)^{1/3}\left(\left(\frac{T}{T_{{\rm F}\uparrow}}\right)^2+\frac{36}{25\pi^2}\left(\frac{T_{{\rm F}\downarrow}}{T_{{\rm F}\uparrow}}\frac{\delta X}{R_\uparrow}\right)^2\right),
\eqname{rel}
\ee
where $\omega_0$ is the oscillation frequency of the order of the harmonic trap frequency, $\delta X$ is the amplitude of the oscillation and $R_\uparrow$ the Thomas-Fermi radius of the majority component. The calculation has been carried out for an homogeneous system and then applied to a trapped gas via the local density approximation. Aside from some numerical factors expression \eq{rel} is typical for the lifetime of quasi-particle in a Fermi liquid \cite{pines}. 
As  can be inferred from \eq{rel} the conditions for reaching the collisionless regime $\omega_0\tau\gg 1$ impose severe conditions from the experimental point of view. Indeed in the experiment \cite{ENScoll}, where the axial compressional mode was investigated in a highly elongated trap, the oscillation of the minority component exhibited strong damping and its frequency could be extracted only through the Fourier transform of the signal. Moreover also the in phase axial breathing mode, which is dominated by the majority component, was found to approach the collisionless value of the frequency only for very high polarization \cite{ENScoll}. 

The collective modes in the collisionless regime can be derived by using the variational principle $\delta S=0$ applied to the action integral
\begin{eqnarray}
S&=&\int dt \langle\Psi|H-i\hbar\partial_t|\Psi\rangle\nonumber\\
&=&\int dt (E-\langle\Psi| i\hbar\partial_t|\Psi\rangle),
\eqname{S}
\end{eqnarray}
where $|\Psi\rangle$ is a multi-parameter many-body wave-function and $E=\langle\Psi|H|\Psi\rangle$ is the energy functional of the system, that we write in the form
\begin{eqnarray}
E=\sum_\sigma\int d{\bf x}\left(\frac{\tau_\sigma}{2m}+\frac{m}{2}(\omega_\perp^2r^2+\omega_z^2 z^2)n_\sigma\right)\nonumber \\+
\frac{3}{5}A\frac{\hbar^2(6\pi^2)^{2/3}}{2m}\int d{\bf x}n_\downarrow n_\uparrow^{2/3}
+a\int d{\bf x}\left(\frac{\tau_\downarrow}{2m}-\frac{n_\downarrow}{2m}\frac{j^2_\uparrow}{n^2_\uparrow}\right),
\eqname{enfunc}
\end{eqnarray}
where $\tau_\sigma/2m$ is the kinetic energy density of the species $\sigma$. The functional \eq{enfunc} accounts for the interaction between the majority and the minority component through the local term in $A$ and the last integral, which is necessary in order to keep into account that the polarons ($\downarrow$-particles) acquires an effective mass due to interaction effects and that galilean invariance implies the presence of a counter current term $j_\uparrow^2$ of the majority component. We include the latter term for completeness, although, since the effective mass is just $10-20\%$ larger than the bare one and the effect of the current term on the frequencies turns out to be fairly small.
Expression \eq{enfunc} corresponds to a typical energy functional to be used in time-dependent Hartree-Fock approaches in the context of small amplitude and low energy oscillations. This approach is equivalent to Landau's theory of Fermi liquid.

At equilibrium the kinetic energy density in Eq. \eq{enfunc} reduces to $\tau_\sigma=\hbar^2(6\pi^2n_\sigma)^{2/3}n_\sigma$ and $j_\uparrow=0$ and, thus, the energy functional Eq. \eq{enfunc} can be used to calculate the density profiles using standard variational procedures. The calculation at equilibrium shows that at very high polarization the majority component is scarcely affected by the interaction and, in paricular, its radius is in practice given by the ideal gas value $R_{\uparrow,i}^0=(48 N_\uparrow)^{1/6}\sqrt{\hbar \bar{\omega}/m\omega_{i}}$, $i=x,\;y,\;z$ with $\bar{\omega}^3=\omega_x\omega_y\omega_z$. Instead the radius of the minority component is quenched with respect to the non interacting gas due to the attractive nature of the force. By taking a Thomas-Fermi description for the minority component (which holds with good accuracy for a large class of experimentally available configurations), one finds that the radius of the minority component is given by the simple expression 
\be
R_\downarrow/R_\uparrow^0=\left(\frac{1-P}{1+P}\right)^{1/6}\left(\left(1+{3 \over 5}A\right){m^* \over m}\right)^{-1/4}.
\eqname{radii}
\ee  
From Eq. \eq{radii} it is seen that the minority radius is quite flat as a function of $P$ except at very high polarization when it goes to zero. We find that this behaviour is reflected in the collective mode frequencies (see below).

In order to determine the solution of the time-dependent problem we develop a variational approachbased on the scaling transformation
\be
\psi_{\sigma}(r,z,t)={\text e}^{-1/2(2\as+\bs)}\psi_{\sigma}^0({\text e}^{-\as}r,{\text e}^{-\bs}z){\text e}^{i(\ks r^2+\xs z^2)}.
\eqname{scaling}
\ee
applied to the single particle wave-functions $\psi_\sigma$ of the two spin species $\sigma=\uparrow,\;\downarrow$, and where $r^2=x^2+y^2$ and $z$ are the radial and the axial coordinate, respectively.
The scaling transformation depends on $4+4$ time-dependent parameters and the corresponding equations are obtained by imposing the variation $\delta S=0$ with $S$ as in \eq{S} with $\langle\Psi|H|\Psi\rangle$ given by Eq. \eq{enfunc}. With respect to a typical hydrodynamic energy functional, Eq. \eq{enfunc} accounts for the deformation of the Fermi surface produced by the scaling ansatz. This effect arises from the kinetic energy density term $\tau_\sigma$ and exploits the elastic nature exhibited by a Fermi liquid in the collisionless regime. The use of hydrodynamic theory would actually yield wrong predictions for the oscillations of these Fermi gases.

By making the choice \eq{scaling} for the scaling transformation we implicitely assume that the equilibrium configuration is axially symmetric and restrict our study to the coupled surface compressional modes of axial/radial nature. 
The reason why we restrict the analysis to such modes is related to the fact that the easiest experimental way of exciting the spin modes in trapped Fermi gases is through a sudden change of the value of scattering length (see Appendix A for the single polaron case). This procedure, which mainly affects the motion of the impurities, is not able to excite other important oscillations like, e.g., the dipole mode of the minority component and it is the actual procedure employed in the recent experiment reported in Ref. \cite{ENScoll}.

The collective modes are just small oscillations around equilibrium, i.e., solutions of the equations of motion derived from the action expanded to  second order in the scaling parameters.
The first order expansion of the action $S$ with respect to the scaling parameters takes contribution only from the energy functional Eq. \eq{enfunc}. Then, the condition $\delta S=0$ provides a relation between the kinetic and the interaction energy equivalent to the virial theorem. Defining the effective mass as $m/m^*=(1+a)$ and the averages $N_\sigma\langle f\rangle_\sigma=\int f(r,z)n_\sigma(r,z)$, we obtain
\begin{widetext}
\begin{eqnarray}
&&-\frac{4}{3}\int \frac{\tau_\uparrow}{2m}+N_\uparrow m\omega_\perp^2\langle r^2\rangle_\uparrow-
N_\downarrow \frac{\hbar^2 (6\pi^2)^{2/3}}{2m} A\left(\langle r\partial_r n_\uparrow^{2/3}\rangle_\downarrow+\frac{4}{3}\langle n_\uparrow^{2/3}\rangle_\downarrow\right)=0\\
&&-\frac{4}{3}\int \frac{\tau_\downarrow}{2m^*}+N_\downarrow m\omega_\perp\langle r^2\rangle_\downarrow+
N_\downarrow \frac{\hbar^2 (6\pi^2)^{2/3}}{2m} A\langle r\partial_r n_\uparrow^{2/3}\rangle_\downarrow=0
\eqname{virial}
\end{eqnarray}
\end{widetext}
Similar relations hold along $z$. In Eq. \eq{virial} all the densities are to be calculated at equilibrium.
The term in the action which depends on the time derivative of the wave-function does not give rise to linear terms due to time reversal symmetry. The quadratic term is given by
\be
\langle\Psi| i\hbar\partial_t|\Psi\rangle^{(2)}=2\sum_\sigma N_\sigma(\langle r^2\rangle_\sigma\as\dot\xs+\langle z^2\rangle_\sigma\bs\dot\ks).
\ee  
Summing up all the contributions and imposing the variational procedure $\delta S=0$, we get eight coupled equations of motion. Four of them represent continuity equations and are relations between the current parameters $\xs$, $\ks$ and the densiy ones $\as,\bs$. In this way we are left to solve a linear system of $4$ equations. 

We find that the two lowest frequency are almost independent of the ratio $N_\downarrow/N_\uparrow$ and they are very close to the ideal gas values $\omega=2\omega_\perp$ and $\omega=2\omega_z$. We call them the in phase modes since the majority and the minoity components move in phase. The frequencies of the other two modes, that we name out of phase or spin modes, can be written as $\omega=2 C_1 \omega_\perp$ and $\omega=2 C_2 \omega_z$ with the renormalization factors turning out to be very close, i.e., $C_1\simeq C_2$. They basically  correspond to the radial and axial motion of the minority component moving in opposite phase with respect to the majority one. In the limit of a single impurity we recover the values $C_1=C_2=\sqrt{\left(1+3/5 A\right)m/m^*}$ of Eq. \eq{single}. 

\begin{figure}
\begin{center}
\includegraphics[height=5.5cm]{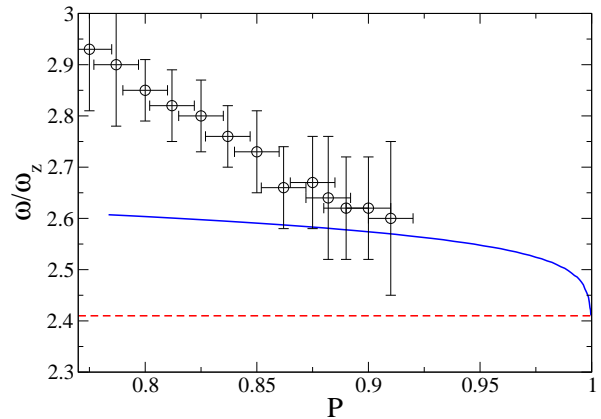}
\caption{Frequency of the axial compressional mode as a function of the polarization $P$. Dashed line: the single polaron mode frequency. Solid line: the collisionless value of the mode frequency obtained via the variational principle described in the present work. Points: experimental data reported in \cite{ENScoll}.}
\label{fig:freq}
\end{center}
\end{figure}

In Fig. (\ref{fig:freq}) we report the result for the axial spin mode as a function of the polarization of the system together with the experimental data of \cite{ENScoll}. We notice that at high polarization the correction to the polaron frequency  Eq.\eq{single}, as a function of the polarization, follows the law $(1-P)^{1/6}$ characterizing the radius of the minority component (see Eq. \eq{radii}).
From Fig. (\ref{fig:freq}) we see that there is a qualitative difference between the present theory and the experiment \cite{ENScoll}, but a remark is due here. Our prediction was derived in the colliosionless regime, while the experiment clearly shows that the in phase mode frequency is strongly affected by collisions even for the highest polarization available. Thus the question of how the observed frequency in Fig. (\ref{fig:freq}) can be compared with our prediction has not an obvious answer \cite{note}.
Our prediction agrees better with the experimental datas at the highest polarization points, where the collisionless approximation is better satisfied \cite{ENScoll}. In this respect we expect that a measurement of the radial compressional mode, should give a much better insight into the problem since in highly elongated traps the radial frequency is much higher than the axial one and thus collisions are less effective. Such an experiment would provide a stringent test for the applicability of Landau theory to the dynamics of these novel systems.

As a last comment, we should remind that the energy functional \eq{enfunc} is strictly valid only at high polarization, being an expansion in the concentration $n_\downarrow/n_\uparrow$ \cite{importance,Loboetal}. It works however quite well when applied to investigate the static properties even till the critical concentration. For instance the inclusion of the next term proportional to $(n_\downarrow/n_\uparrow)^2$ \cite{importance,pilati,chevyx2} introduces only small corrections to the values of the radii, even close to the critical polarization limit $P_C=0.77$. Such a term does not change the picture emerging from the present analysis.

A natural extension of this work would be to study the collective mode frequencies for the impurity out of resonance. In particolar for positive values of the scattering length it would be interesting to study the distinction between the polaron state and the molecular one from a dynamical point of view.

\acknowledgments

We aknowledge N. Navon for the experimental datas and F. Chevy, S. Giorgini, C. Salomon and W. Zwerger for useful discussions.                                                                              
This research has been supported by EuroQUAM Fermix program and by MIUR PRIN 2007.

\appendix

\section{Sudden change of the scattering length and  polaron dynamics}

What does it happen to the dyamics of a single impurity in a Fermi sea if one suddenly change the value of the scattering lenght?

In presence of an harmonic trapping potential $V_{ho}$ we can write for the polaron an effective time dependent Hamiltonian as
\be
H=\sum_{i=1}^3\left({p^2 \over 2m(t)} + {1 \over 2}m(t)\oit^2 x_i^2\right),
\label{eq:Ht}
\ee
where $m(t)$ is the istantaneous value of the effective mass and we define an effective frequency $\omega^2_i(t)=m/m(t)(1+3/5A(t))\oi^2$. The Hamiltonian \eq{Ht}, is easily diagonalized to the form $H=\sum \hbar\oit\ai^\dagger \ai$ by introducing the istantaneous ladder operators 
\bea
\ai&=&{1\over\sqrt{2\hbar}}\left(u_i(t)x_i+i{1\over u_i(t)}p_i\right),\\
\ai^\dagger&=&{1\over\sqrt{2\hbar}}\left(u_i(t)x_i-i{1\over u_i(t)}p_i\right),
\eea
where $u_i(t)=\sqrt{m(t)\oit}$ is fixed by the way  the scattering length is changed in time.
In the Heisenberg picture their evolution is simply
\bea
\dot\ai&=&-i\oit\ai+{{\dot u_i(t)}\over u_i(t)}\ai^\dagger,\\
\dot\ai^\dagger&=&i\oit\ai^\dagger+{{\dot u_i(t)}\over u_i(t)}\ai.
\eea
The first terms on the RHS describes the trivial evolution of the operators, while the second ones take into account the explicit time dependence of the parameters in the Hamiltonian and are responsible for the mixing between $a_i$ and $a^\dagger_i$, which eventually depends on the rate at which the scattering length is varied in time.

Once the functional form of $u_i(t)$ is fixed it is easy to calculate the evolution of the width of the impurity wavefunction $\Delta x_i^2$.
In the case of a sudden change (see also \cite{tdho}) it is easy to write the evolution of the ladder operators as 
\be
\ai={1 \over 2}\left(\left({u_{i,f}\over u_{i,0}}+{u_{i,0}\over u_{i,f}}\right)a_{i,0}+\left({u_{i,f}\over u_{i,0}}-{u_{i,0}\over u_{i,f}}\right)a_{i,0}^\dagger\right)e^{-i\omega_{i f} t}.
\ee
where $\omega_{i f}$ is the final value of the effective trapping frequency, $u_{i,0}$ and $u_{i,f}$ are the initial and the final values of $u_i$, respectively. Let us suppose we start from a situation where we do not have any interaction between the two-hyperfine levels for which $m(0^-)=m$ and $A(0^-)=0$, for which $u_{i,0}=\sqrt{m\omega_i}$, and the system is in its ground state. Suddenly we change the scattering to a very large value (unitarity), for which $m(0^+)=m^*$ and $A(0^+)=A$, i.e., $
u_{i,f}=\sqrt{\omega_i}(m m^* (1+3/5 A))^{1/4}$.
In this case the width of the polaron along the {\it i}-th direction will oscillate at a frequency given by twice the oscillator frequency \eq{single}  around the new equilibrium position $\Delta x_i^2=1/(4 m\oi)\left(1+(m^*/m(1+3/5 A)^{-1}\right)$ according to
\begin{widetext}
\be
\Delta x_i^2(t)={1\over {4 m\oi}}\left[\left(1+{1\over{m^*/m(1+3/5 A)}}\right)+\left(1-{1\over{m^*/m(1+3/5 A)}}\right)\cos\left(2\omega_i\sqrt{{m\over m^*}(1+3/5 A)}t\right)\right].
\ee 
\end{widetext}
Thus, the measurement of the  frequency of the oscillation and of its amplitude provides the values $A$ and $m^*$ of the parameters of the polaron.


\begin{thebibliography}{99}
\bibitem{note}
Undestanding the discrepancy between theory and  experiment in terms of collisional effects is not however obvious. In fact collisions usually reduce the value of the frequencies with respect to their collisionless values.
\end{thebibliography}
\end{document}